\documentclass{article}
\usepackage{spconf,amsmath,graphicx}
\usepackage{subcaption}
\expandafter\def\csname ver@subfig.sty\endcsname{}

\usepackage{svg}
\usepackage{hyperref}
\usepackage{enumitem}
\usepackage{float}
\setlist{nosep, leftmargin=14pt}
\usepackage{subcaption}
\usepackage{mwe} % to get dummy images
\usepackage{multirow}
\usepackage{hyperref}
\usepackage{xcolor}
\usepackage{booktabs}
\usepackage{graphicx}
\usepackage{textcomp}
\usepackage{tabularx}
\usepackage{mathrsfs}
\usepackage{mathtools}
\usepackage[normalem]{ulem}
\usepackage{subfig}
\usepackage{tabularx}
\RequirePackage[numbers,sort&compress]{natbib}
\title{Semise: Semi-supervised learning for severity representation in medical image}
\name{Dung T. Tran$^{\star}$ \qquad 
Hung Vu$^{\dagger \dagger}$ \qquad 
Anh Tran$^{\star}$ \qquad 
Hieu Pham$^{\dagger \dagger}$ \qquad 
Hong Nguyen$^{\ddagger}$ \qquad 
Phong Nguyen$^{\star}$}
\address{$^{\star}$ Hanoi University of Science and Technology, Hanoi, Vietnam \\
 $^{\ddagger}$ University of Southern California, Los Angeles, United States \\
 $^{\dagger \dagger}$ College of Engineering \& Computer Science, VinUniversity, Hanoi, Vietnam\\
 $^{\dagger \dagger}$ VinUni-Illinois Smart Health Center, VinUniversity, Hanoi, Vietnam }

\begin{document} 
\maketitle

\begin{abstract}
This paper introduces \textbf{SEMISE}, a novel method for representation learning in medical imaging that combines self-supervised and supervised learning. By leveraging both labeled and augmented data, \textbf{SEMISE} addresses the challenge of data scarcity and enhances the encoder’s ability to extract meaningful features. This integrated approach leads to more informative representations, improving performance on downstream tasks. As result, our approach achieved a 12\% improvement in classification and a 3\% improvement in segmentation, outperforming existing methods. These results demonstrate the potential of SIMESE to advance medical image analysis and offer more accurate solutions for healthcare applications, particularly in contexts where labeled data is limited.
\end{abstract}
\vspace{-10pt}
\section{Introduction}
\label{sec:intro}
\vspace{-5pt}

Severity study \cite{kollias2022aimiacovid19detection, KALPATHYCRAMER20162345} in medical domain is critical problems in which medical experts identify severity of illness of subjects by different levels (mild, moderate, severe, extremely severe)  \cite{10.1016/j.cmpb.2022.106947, Li2020-ht, Gu2023-dq, qiblawey2021detectionseverityclassificationcovid19, Tran2022ANT, Yildiz2019-np} or preferred comparison (one is more severe than the other) \cite{nguyen2024conprolearningseverityrepresentation}. 
% It is nessessary to evaluate severity of pathology
Investigating severity can help putting hospital in order by clustering patients have same level of illness or better queue for timely treatment.
% As a example, World had seen coronavirus disease 2019 globally overcrowded hospitals \cite{qiblawey2021detectionseverityclassificationcovid19} and cause significant mortality. 
Several severity study \cite{Lauque2022-nr, Azadeh-Fard2016-by} have been conducted to relate severity level to length of hospital stay or inpatient mortality. 
Yet, taking doctors to rank thousand to millions of patients' reports is either time consuming or economic inefficient.
This paper aim to develop an AI model to learn severity knowledge in medical images and generalize well on multiple tasks including severity classification and pathology segmentation. We put our focus on images although there is more than one medium to identify severity of pathologies, such as bioSignal \cite{Ancillon2022-gt}, EEG \cite{Tesh2022-sr}, etc.

% Severity assessment in the medical field is essential for identifying a patient's condition, but it remains challenging due to the time and cost involved in expert evaluations [cite]. To mitigate this, several machine learning models have been developed to study severity in various perspective. Recently, representation learning has become a key component of advanced AI systems [cite], enabling meaningful features extraction re contextual domain knowledge from unlabled and labled data. Unlike traditional machine learning methods, which rely on manually crafted features—an expensive and often impractical approach in data-limited medical domains—self-supervised and semi-supervised methods leverage large amounts of unlabeled data. This enables the learning of domain-specific representations, improving robustness and generalization across multiple downstream tasks, such as severity classification, injury detection, or segmentation.

% With the increasing importance of visual representation learning, four primary approaches have gained attention:
Recently, representation learning has emerged as a foundational component in any advance AI systems, as it facilitate the automatic features extraction to learn signature contextual information from data.
In literature, representation learning can be categorized into Self-Supervise, Supervised and Semi-Supervised Learning.
% \noindent
% \textbf{Self-Supervised Representation Learning (SSL):} Self-Supervised Learning in literature can be categorized into contrastive and generative models. Several work have been working on Contrastive SSL, including but not limit to SimCLR \cite{chen2020simpleframeworkcontrastivelearning}, SimMIM \cite{xie2022simmimsimpleframeworkmasked}, MoCo \cite{he2020momentumcontrastunsupervisedvisual}, SwAV \cite{caron2021unsupervisedlearningvisualfeatures}, DINO \cite{caron2021emergingpropertiesselfsupervisedvision}. The under-mechanism is to improves local understanding of embedding by maximizing similarity between augmented views of the same image. Generative representation learning, such as MAE \cite{he2021maskedautoencodersscalablevision}, SODA \cite{hudson2023sodabottleneckdiffusionmodels}, DiffAE \cite{preechakul2022diffusionautoencodersmeaningfuldecodable}, embeds enough visual knowledge to low-dimension vector so that it can regenerate original image from its noisy version.

\noindent
\textit{Self-Supervised Representaion Learning (SSL): } SSL methods, categorized into contrastive and generative models. Contrastive methods \cite{chen2020simpleframeworkcontrastivelearning, xie2022simmimsimpleframeworkmasked, he2020momentumcontrastunsupervisedvisual,caron2021unsupervisedlearningvisualfeatures,caron2021emergingpropertiesselfsupervisedvision}, enhance local embedding understanding by maximizing similarity between augmented views of the same image while generative models  \cite{he2021maskedautoencodersscalablevision,hudson2023sodabottleneckdiffusionmodels,preechakul2022diffusionautoencodersmeaningfuldecodable} focus on embedding visual knowledge into low-dimensional vectors to regenerate original images from noisy versions. 
% However, effective evaluation metrics remain a challenge.

% \noindent
% \textbf{Supervised Representation Learning (SRL):} While SSL learn local information, SRL inject cross-subject information SupCon utilizes supervised contrastive loss with class labels. Several work such as SERL \cite{luo2024serlsoftwaresuitesampleefficient}, SupCon \cite{khosla2021supervisedcontrastivelearning} show promissing results on downstream task.  however, they all relies heavily on large labeled datasets, which are often scarce in medical contexts.

\noindent
\textit{Supervised Represenation Learning (SRL):} SRL models such as SupCon \cite{khosla2021supervisedcontrastivelearning} and SERL \cite{luo2024serlsoftwaresuitesampleefficient} utilize labeled datasets to incorporate cross-subject information, yielding promising results in downstream tasks. 
% However, these approaches depend heavily on large labeled datasets, which are often scarce in medical contexts.
However, these approaches construct embedding space base on similarity measure but not relative relation between classes, e.g in medical context, mild severity class close to normality than extreme severity. Recently, ConPro \cite{nguyen2024conprolearningseverityrepresentation} employ preference comparisons to rank disease severity, optimizing latent embeddings based on preference information. However, these models often rely on limited severity labels and focus primarily on classification tasks. This study aims to extend these approaches to broader downstream tasks while addressing their limitations.
\begin{figure*}[h]
    \centering
    \vspace{-4em}
    \includegraphics[width=\linewidth]{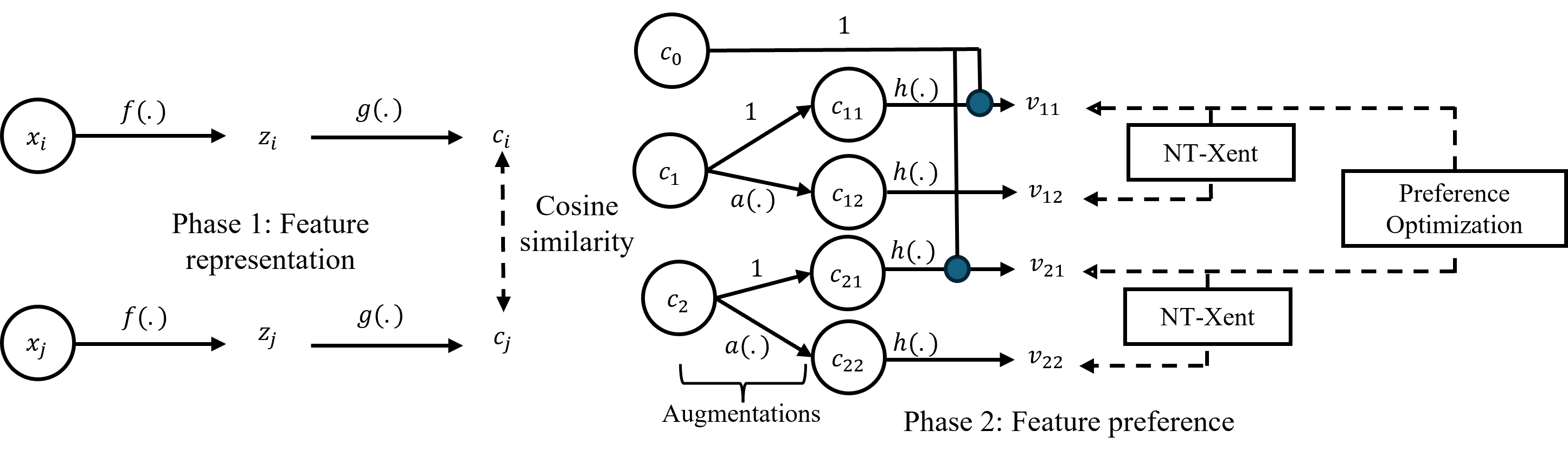}
    \vspace{-2.5em}
    \caption{SEMISE learning framework}
    \vspace{-1.5em}
    \label{fig:ConSimPro-learning-framework}
\end{figure*}

\noindent
\textit{Semi-Supervised Representation Learning (SemiSL):} SemiSL combines labeled and unlabeled data, with frameworks like ROPAWS\cite{mo2023ropawsrobustsemisupervisedrepresentation}, FixMatch \cite{sohn2020fixmatchsimplifyingsemisupervisedlearning}, and Mean Teacher \cite{tarvainen2018meanteachersbetterrole} addressing semi-supervised task challenges. Despite their potential, these methods face issues related to unreliable pseudo-labels and class imbalance. The proposed approach leverages SemiSL to integrate the strengths of the aforementioned techniques.
% \\
% Despite recent advancements, optimizing the influence of upstream tasks, such as feature extraction and noise reduction, on downstream tasks like segmentation and diagnosis in medical imaging remains challenging. Our work enhances the encoder block's performance, with contributions including:
% This hybrid technique has demonstrated improved robustness and accuracy across various medical imaging datasets.
% }

To address mentioned challenges, we summarized our contribution as follow
\begin{itemize}
    \item We proposed \textbf{SemiSe}: A \textbf{Semi}-supervised learning framework that leverages both in-context self-learning and cross-subject label information to optimally enrich \textbf{Se}verity knowledge into latent embedding vectors. %including severity classification and pathology-specific segmentation
    \item We quantitatively evaluate our framework on multiple downstream tasks and show that our proposed framework outperform predecessors in term of F1 score, MAEE, IoUs and DICE.
\end{itemize}
\vspace{-10pt}
\section{Methodology}
\label{sec:pagestyle}
\vspace{-5pt}
The proposed methodology, SemiSe as detailed in \cite{nguyen2024conprolearningseverityrepresentation} and depicted in Fig. \ref{fig:ConSimPro-learning-framework}, comprises two main phases. 
The initial phase discriminates Healthy-Anomalie, while the subsequent phase centers around preference optimization. 
In the latter phase, self-supervised learning techniques, particularly SimCLR, are incorporated along with reference opt to both learn the in-context and cross-subject representation.
% During the initial phase, our goal is to maximize the latent distance between the normal and abnormal classes. In the subsequent phase, we apply the Preference Optimization function outlined in \cite{nguyen2024conprolearningseverityrepresentation}, in conjunction with SimCLR \cite{chen2020simpleframeworkcontrastivelearning}, to effectively rank the severity levels of abnormalities based on reference vectors derived from the normal class.

\begin{table*}[t]
\centering
\setlength{\tabcolsep}{8pt}
\caption{Multiclass classification results.}
\label{tab1}
\begin{tabular}{lccccccccc}
\hline
 &  \multicolumn{3}{c}{VinDr-Mammo} & \multicolumn{3}{c}{Papilledema} & \multicolumn{3}{c}{ISIC}\\
\cmidrule(r){2-4} \cmidrule(r){5-7} \cmidrule(r){8-10} 
{Methods}&F1 & MAEE &Recall & F1 &MAEE & Recall & F1 &MAEE & Recall\\
\hline
SODA  &16.10 & 3.04 & 20,00 & 43,90 & 1.37 & 50,00 & 10,70 & 58,51 & 14,30 \\
SupCon-2  & 17.70 & 2.94 & 20,20 & 24.16 & 2.74 & 33.30 & \underline{56.38} & \underline{27.94} & 53.46\\
SupCon-n  & 18.80 & 2.93 & 20,30 & 24.16 & 2.74 & 33.30 & 55.72 & 29.69 & \underline{54.58}\\
SimCLR & 17.64 & 2.96 & 20.10 & 34.24 & 2.32 & 40.26 & 54.62 & 28.08 & 53.52\\
ConPrO & \underline{20.50} & \underline{2.60} & \underline{21.70} & \underline{94.10} & \underline{1.13} & \underline{93.76} & 55.92 & 28.11 & 54.14\\[0.5ex]
% \vspace{3pt}
\hline
% \textbf{Ours  ($\alpha = 0.1)$ } & \textbf{20.40} & \textbf{2.65} & \textbf{21.80} & \textbf{94.80} & \textbf{1.06} & \textbf{93.00} & \textbf{55.20} & \textbf{29.43} & \textbf{52.20} \\
% \textbf{Ours  ($\alpha = 0.2)$ } & \textbf{20.50} & \textbf{2.63} & \textbf{21.90} & \textbf{94.80} & \textbf{1.06} & \textbf{93.00} & \textbf{57.90} & \textbf{25.47} & \textbf{54.90} \\
% \textbf{Ours  ($\alpha = 0.3)$ } & \textbf{20.40} & \textbf{2.63} & \textbf{21.80} & \textbf{94.80} & \textbf{1.06} & \textbf{93.00} & \textbf{59.10} & \textbf{27.56} & \textbf{54.90} \\
% \textbf{Ours  ($\alpha = 0.4)$ } & \textbf{23.80} & \textbf{2.59} & \textbf{23.50} & \textbf{94.80} & \textbf{1.06} & \textbf{93.00} & \textbf{58.10} & \textbf{29.30} & \textbf{54.40} \\
\textbf{SEMISE  ($\alpha = 0.5)$ } & \textbf{32.34} & \textbf{2.57} & \textbf{29.60} & \textbf{94.84} & \textbf{1.13} & \textbf{94.56} & {55.10} & {31.90} & {52.40} \\
% \textbf{Ours  ($\alpha = 0.6)$ } & \textbf{26.80} & \textbf{2.61} & \textbf{26.10} & \textbf{94.80} & \textbf{1.07} & \textbf{94.00} & \textbf{56.70} & \textbf{25.82} & \textbf{54.90} \\
\textbf{SEMISE  ($\alpha = 0.7)$ } & {26.80} & {2.61} & {26.10} & {94.80} & {1.06} & {93.00} & \textbf{59.40} & \textbf{26.80} & \textbf{56.60} \\
% \textbf{Ours  ($\alpha = 0.8)$ } & \textbf{26.80} & \textbf{2.62} & \textbf{26.10} & \textbf{94.80} & \textbf{1.06} & \textbf{93.00} & \textbf{55.70} & \textbf{27.37} & \textbf{53.10} \\
% \textbf{Ours  ($\alpha = 0.9)$ } & \textbf{26.80} & \textbf{2.62} & \textbf{26.10} & \textbf{94.00} & \textbf{1.07} & \textbf{91.90} & \textbf{-} & \textbf{-} & \textbf{-} \\
\hline
\end{tabular}
    \label{tab:table-1}
\end{table*}

\begin{table*}[ht]
\begin{minipage}[ht]{0.3\linewidth}
\centering
\begin{tabular}{lccccccccc}
\toprule
     \textbf{Methods} & \textbf{IoU (\%)} & \textbf{DICE (\%)} \\
    \hline
    SODA & 28.09 & 39.80 \\
    SimCLR & 41,62 & 56,58 \\
    SupCon-2 & 41,54  & 56,85  \\
    SupCon-n & 40,68  & 55,71  \\
    ConPro & \underline{42,22}  & \underline{57,17}  \\
    \textbf{SEMISE} & \textbf{43,10 } & \textbf{58,06} \\
\bottomrule
\end{tabular}
\caption{Segmentation Results on the ISIC Dataset.}
\label{table:segmentation-sheet}
\end{minipage}\hfill
\begin{minipage}[ht]{0.4\linewidth}
\centering
\includegraphics[width=\linewidth] {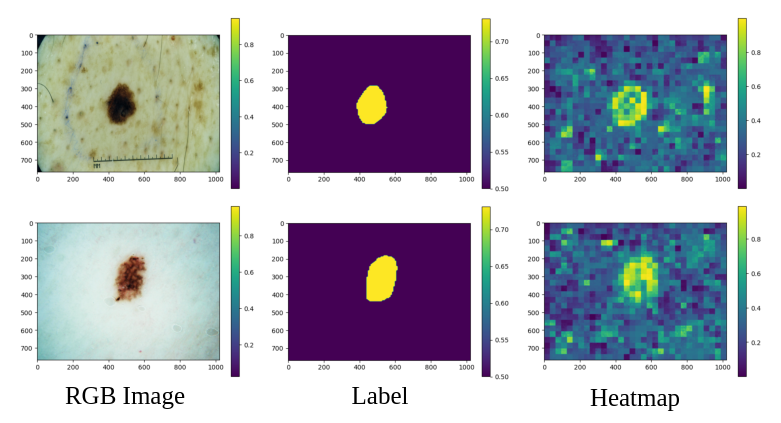}
\captionof{figure}{Heatmap of the segmentation results.}
\label{fig:segmentation-output}
% \end{minipage}
\end{minipage}\hfill
\begin{minipage}[ht]{0.29\linewidth}
\centering
\includegraphics[width=1\linewidth] {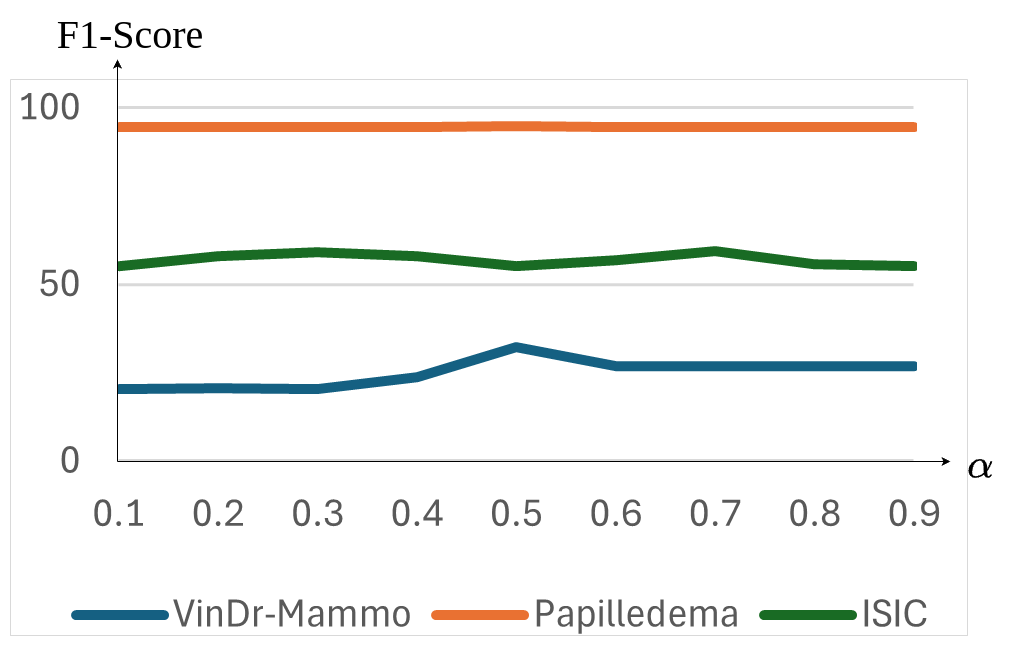}
\captionof{figure}{Correlation between $\alpha$ and F1-Score across Three Datasets.}
\label{fig:alpha-f1}
\end{minipage}
\end{table*}
\vspace{-10pt}
\subsection{Healthy-Anomaly Discrimination}
To classify a sample as healthy or anomalous, it is processed through a feature extraction block \( f(.) \) and a projection block \( g(.) \) to create its representation. Utilizing contrastive loss \cite{khosla2021supervisedcontrastivelearning} with labeled data enhances the feature extraction process, enabling the model to distinguish the unique characteristics of healthy samples from anomalies. The contrastive loss function can be defined as follows:
\vspace{-10pt}
\begin{equation}
\mathcal{L}_{SupCon} = \frac{1}{2N} \sum_{i=1}^{N} \left[ y_i D^2 + (1 - y_i) \max(0, m - D)^2 \right],
\label{loss-supcon}
\end{equation}
where \( N \) is the total number of sample pairs, \( y_i \) indicates whether the samples are similar (1) or dissimilar (0), \( D \) represents the distance between the representations of the samples (with cosine distance used in this work), and \( m \) is the margin that separates dissimilar pairs.
\vspace{-10pt}
\subsection{Contrastive Representation Learning}
\label{ssec:subhead}
After distinguishing between healthy and anomalous samples, self-supervised contrastive learning allows the model to examine multiple perspectives of an anomalous sample by contrasting the original with its augmented counterpart, using the augmentation function \( a(\cdot) \) and projection head \( h(\cdot) \). This process encourages the formation of distinct clusters of similar features in the latent space. To achieve this, the NT-Xent loss \cite{agren2022ntxentlossupperbound} from the SimCLR framework \cite{chen2020simpleframeworkcontrastivelearning} is utilized:
\begin{equation}
\mathcal{L}_{\text{NT-Xent}} = \frac{-1}{2N} \sum_{i=1}^{N} \left[ \log \frac{\exp(\text{sim}(z_i, z_j) / \tau)}{\sum_{k=1}^{2N} \textbf{1}_{[k \neq i]} \exp(\text{sim}(z_i, z_k) / \tau)} \right],
\label{Ntxentloss}
\end{equation}
where $z_i$ and $z_j$ are the feature representations of the original sample and its augmented version, respectively, \text{sim}($z_i$, $z_j$) denotes the cosine similarity between $z_i$ and $z_j$, $\tau$ is a temperature scalar that helps control the sharpness of the distribution and $N$ typically refers to the number of examples in a mini-batch.

% \begin{itemize}
%     \item \( z_i \) and \( z_j \) are the feature representations of the original sample and its augmented version.
%     \item \( \text{sim}(z_i, z_j) \) denotes the cosine similarity between \( z_i \) and \( z_j \).
%     \item \( \tau \) is the temperature parameter.
%     \item \( N \) is the number of positive pairs.
%     \item \( \textbf{1}_{[k \neq i]} \) is an indicator function that equals 1 if \( k \neq i \) and 0 otherwise.
% \end{itemize}
\vspace{-10pt}
\subsection{Preference Optimization}
Preference optimization enhances the representations of samples in latent space, using a healthy sample as a reference. Anomalous samples are evaluated as pairs, with the more severe sample positioned further from the reference. We modified the Preference Comparison Loss \cite{nguyen2024conprolearningseverityrepresentation} to better address our specific issue. Instead of using negative log-likelihood loss for re-parameterizing the feature space, we opted for binary cross-entropy loss (BCE), as it helps the model converge faster in our experiments. The modified loss function is defined as follows:
\begin{equation*}
\hspace{-1.5in}
\mathcal{L}_{PrO} (\nu_i, \nu_j, \pi_0, y_{ij} \mid r^* = d_{\cos})
\end{equation*}
\begin{equation}
= \text{BCE} \left[ \log \left( \sigma \left( r^* (\nu_i, \pi_0) - r^* (\nu_j, \pi_0) \right) \right) \right],
\label{eq:Pro-los}
\end{equation}
in which $\pi_0$ the reference feature vector. The goal is to assess the distances from \(\nu_i\) and \(\nu_j\) to \(\pi_0\). If \(\nu_i\) is more dangerous than \(\nu_j\) (label 1), \(\nu_i\) should be further from \(\pi_0\) than \(\nu_j\), and vice versa.
\vspace{-5pt}
\subsection{Combined Loss Function between SSL and SRL}
To learn domain knowledge and cross-subject correlation simultaneously, the objective functions of SSL and SRL are combined to enhance model performance. The two previously described loss functions (\ref{Ntxentloss}), (\ref{eq:Pro-los}) are combined into a unified loss function. This function includes a weighting parameter \(\alpha\) to adjust the relative importance of each component:
\vspace{-5pt}
\begin{equation}
\mathcal{L}_{\text{combine}} = \alpha \cdot \mathcal{L}_{\text{NT-Xent}} + (1 - \alpha) \cdot \mathcal{L}_{PrO}
% \vspace{-2em}
\end{equation}

% Here, \(\alpha\) is a hyperparameter that balances the two losses. A default value of 0.5 is used in experiments, indicating equal importance for both components, though it can be a learnable parametter while evaluating.
% By combining the loss functions, the goal is to leverage their strengths for a more robust model.
\vspace{-15pt}
\section{Experimental Settings}
\vspace{-5pt}
\subsection{Datasets}
% \begin{table}[h]
%   \begin{center}
%     \caption{Statistics of three datasets.}
%       \label{tab:dataset}
%     % \small % Use smaller font size to help with width
%     \begin{tabular}{l|ccp{2.5cm}} % Use p{3cm} for better text wrapping
%       \hline
%       \textbf{Dataset} & \textbf{Train} & \textbf{Test} & \textbf{Task} \\
%       \hline
%       VinDr-Mammo & 4000 & 1000 & Class labels, bounding-box annotations \\ 
%       Papilledema  & 500+  & 100+  & Class labels \\ 
%       ISIC         & 20000+ & 5000+ & Class labels, Segments Annotation \\ 
%       \hline
%     \end{tabular}
%   \end{center}
%   % \vspace{-2em}
% \end{table}
% Table \ref{tab:dataset} summarizes key statistics, including dataset names, training sizes, test sizes, and available annotations.

\noindent
\textit{VinDr-Mammography:} 
    The VinDr-Mammography dataset \cite{nguyen2023vindrmammolargescalebenchmarkdataset} consists of over 5,000 annotated mammograms labeled by radiologists. It includes various abnormalities and has 5 levels of severity. The dataset also includes bounding-box annotation for lesions, such as masses and calcification's.

\noindent
\textit{Papilledema} 
 is Kaggle public dataset in which the pathology
 \cite{avramidis2022automatingdetectionpapilledemapediatric} is presented by the swelling of the optic disc due to increased intracranial pressure, indicating serious neurological conditions. Without loss of generality, we assume that the condition can be classified into three severity levels: normal, pseudo-papilledema, and papilledema. Among these, normal presents the least risk, while papilledema is the most severe, requiring immediate medical attention.

\noindent
\textit{ISIC Skin Lesion} 
The ISIC skin lesion dataset \cite{gutman2016skinlesionanalysismelanoma} is a comprehensive collection of dermoscopic images for melanoma and skin lesion classification, annotated by dermatologists. It supports research in skin cancer detection, segmentation, and classification.

% \subsection{Downstream Tasks}
% \label{sec:typestyle}
% \textcolor{blue}{The same encoder is used for feature extraction across all tasks. Detailed implementations are below.}

% \noindent
% \textit{Severity Classification:} 
% \textcolor{blue}{A simple architecture with two dense ReLU-activated layers and a Dropout layer (0.3 probability) follows the ResNet embedding to mitigate overfitting. Only the fully connected layer is trained, while the ResNet encoder remains frozen.}

% \noindent
% \textit{Pathology Segmentation:} 
% \textcolor{blue}{A pretrained encoder extracts features from input images, processed by a segmentation network to identify regions of interest. The encoder is frozen, and the decoder is fine-tuned using a reused UNet architecture for pixel-level segmentation.}
\vspace{-10pt}
\subsection{Experimental Setup.}
All experiments used the same ResNet-50 encoder architecture for upstream task and downstream tasks such as classification and segmentation task, conducted on a GTX 4070 with a batch size of 256. 
Datasets were split to prevent leakage: 72/8/20 for VinDr-Mammo, 70/15/15 for Papilledema and ISIC dataset. For preference optimization, \(10^5\) pairs were randomly selected for training and \(10^3\) pairs were chosen for evaluation. SGD with a momentum of 0.9 was used, updating the encoder with a learning rate of \(10^{-3}\) and the projection heads at \(10^{-1}\). The ResNet-50 outputs 2048-dimensional vectors. The projection head \(g(.)\) is a fully connected layer with a 256-dimensional output, and the preference comparison head \(h(.)\) maintains normality in the preference vector dimensions.
% \begin{itemize}
%     \item \textit{Data Splitting:} Datasets were split to prevent leakage: 70/15/15 for Papilledema, 72/8/20 for VinDr-Mammo, and 70/15/15 for the ISIC dataset challenge. For preference optimization, \(10^5\) pairs were randomly selected for training and \(10^3\) pairs were chosen for evaluation.

%     \item \textit{Optimizer:} SGD with a momentum of 0.9 was used, updating the encoder with a learning rate of \(10^{-3}\) and the projection heads at \(10^{-1}\).

%     \item \textit{Projection Heads:} The ResNet-50 outputs 2048-dimensional vectors. The projection head \(g(.)\) is a fully connected layer with a 256-dimensional output, and the preference comparison head \(h(.)\) maintains normality in the preference vector dimensions.
% \end{itemize}

\noindent
 Without loss of generality, we select representative models SODA \cite{hudson2023sodabottleneckdiffusionmodels}, SimCLR \cite{chen2020simpleframeworkcontrastivelearning}, SupCon-n \cite{khosla2021supervisedcontrastivelearning} and ConPro \cite{nguyen2024conprolearningseverityrepresentation}. for comparion.
% \vspace{-5pt}
%  \subsection{Metrics: } 
 To evaluate the methods, performance is assessed through downstream tasks, including classification and segmentation. For classification, F1 Score, Recall, and MAEE are used to evaluate the effectiveness of our model. In segmentation, Intersection over Union (IoU) and Dice Coefficient are employed, where higher values reflect improved segmentation accuracy. 
 % Overall, these metrics are chosen to comprehensively assess the precision and accuracy of the methods across tasks.
\vspace{-15pt}
\section{Results}
\label{sec:majhead}
\vspace{-5pt}
\subsection{Severity Classification}
A simple architecture with two dense ReLU-activated layers and a Dropout layer (0.3 probability) follows the ResNet embedding to mitigate overfitting. Only the fully connected layer is trained, while the ResNet encoder remains frozen.
These results which is shown in Table \ref{tab:table-1} demonstrate the model's effectiveness in feature extraction, leading to significant classification improvements. The enhancements in F1 Score and Recall show that our approach addresses the limitations of methods like ConPro and SODA, establishing it as a robust solution for medical image classification. Our model is superior than others because the pre-trained encoder can learn features from multiple views of the same sample while distinguishing key features in paired images, resulting in more accurate and robust representations. Fig \ref{fig:alpha-f1} illustrates the effect of the alpha parameter on classification performance through the F1-Score metric.

%\vspace{-5}
\subsection{Segmentation Task}
A pretrained encoder extracts features from input images, processed by a segmentation network to identify regions of interest. The encoder is frozen, and the decoder is fine-tuned using a reused UNet architecture for pixel-level segmentation. The same encoder is used for feature extraction across all tasks.
% \label{ssec:subhead}
% \begin{figure}
%     \centering
%     % \vspace{-0.5em}
%     \includegraphics[width=\linewidth]{Image/Unet.drawio.png}
%     % \vspace{-3em}
%     \caption{U-Net architecture}
%     \vspace{-1em}
%     \label{fig:unet_architecture}
% \end{figure}
Results of segmentation task are shown in Table \ref{table:segmentation-sheet}. Our segmentation model is designed based on the U-Net architecture. After freezing the encoder block and constructing the segmentation model, we evaluated its performance on the ISIC dataset. Figure \ref{fig:segmentation-output} presents the heatmaps resulting from the segmentation outputs of this model. These heatmaps provide a quantitative visualization of the model's effectiveness in accurately delineating significant features within medical images, underscoring its capacity for robust feature identification and spatial awareness.
\vspace{-10pt}
\section{Conclusions}
\vspace{-10pt}
This paper introduces a novel representation learning approach that enhances upstream tasks through a new feature extraction method. The proposed method demonstrates significant improvements in both classification and segmentation tasks, outperforming techniques SOTA representation learning methods across various medical imaging datasets. Key metrics, including F1 Score, IoU, and Dice Coefficient, highlight the robustness of the approach. These results confirm that strengthening upstream representation learning can substantially enhance downstream task performance. 
% Future work will focus on refining this approach and extending its application to additional medical imaging tasks and datasets. Moreover, selecting the hyperparameter $\alpha$ remains a challenge, suggesting the need for an effective algorithm or method to optimize this parameter.

\vspace{-10pt}

\bibliographystyle{unsrt}
\bibliography{refs}

\end{document}